\documentstyle[aps]{revtex}

\newcommand{\veck}{{\vec{k}}}
\newcommand{\vecr}{{\vec{r}}}
\newcommand{\vecrp}{{\vec{r}}{\,}^\prime}
\newcommand{\vecrrp}{{\vec{r}}{\,}^{\prime\prime}}

\begin{document}

\title{Fast Algorithms For Josephson Junction Arrays : Bus--bars
and Defects }
\author{Sujay Datta, Shantilal Das, and Deshdeep Sahdev
\footnotemark[1]
\footnotetext[1]{Email: ds@iitk.ernet.in}}
\address{
	Department of Physics, Indian Institute of Technology, Kanpur 208~016,
	INDIA
}
\author{Ravi Mehrotra
\footnotemark[2]
\footnotetext[2]{Email: ravi@csnpl.ren.nic.in }}
\address{
	National Physical Laboratory, Dr. K.\ S.\ Krishnan Rd., New Delhi 110~012,
	INDIA
}
\author{Subodh R. Shenoy
\footnotemark[3]
\footnotetext[3]{On leave from
{\it School Of Physics, Central University of Hyderabad,  Hyderabad
500~134, INDIA}}}
\address{I.C.T.P., P.O. Box 586, 34100, Trieste, Italy}
\date{\today}
\maketitle

\begin{abstract}
      We critically  review the fast algorithms for the numerical  study
      of  two--dimensional  Josephson  junction  arrays and develop  the
      analogy  of such  systems  with  electrostatics.  We extend  these
      procedures  to arrays with  bus--bars  and  defects in the form of
      missing  bonds.  The role of boundaries and of the guage choice in
      determing  the Green's  function of the system is  clarified.  The
      extension of the Green's function  approach to other situations is
      also discussed.
\end{abstract}


\section{Introduction}
\label{sec:intro}

  The  dynamical  properties  of Josephson  junction  arrays  (JJAs) are
currently   the   foci   of   several   experimental   and   theoretical
investigations   \cite{Nato}.  These   arrays   can  now  be   routinely
fabricated in several sizes and geometries, and the  characteristics  of
their  junctions  can be  varied  at will  over a wide  range of  values
\cite{ZantPRL1}.  A large body of high--precision  experimental data has
consequently  become  available  for JJAs in the  presence  of  external
magnetic fields  \cite{Rzch,Tighe,ZantJLTP}.  On the theoretical  front,
several  insights  into the  behaviour of JJAs have come from  numerical
studies  of  the  underlying   equations  of  motion  as  given  by  the
resistively- and capacitively-shunted  junction (RCSJ) model using input
current  drives  and  defects,  both  controlled  \cite{Xia}  and random
\cite{Doming}.  With  the  size  of   experimental   arrays   increasing
continuously, and with the number of interesting  effects best seen only
in large arrays going up in equal  measure, it has become  imperative to
find ever more efficient  algorithms for implementing the  corresponding
simulations,  inclusive of all the experimental  conditions.  An example
of the latter for  current--driven  arrays is the  presence of bus-bars,
through  which the  external  current can be  conveniently  injected  or
withdrawn.

   To  understand  the problem which these  algorithms  must address, we
recall that in the RCSJ model, the total current,  $i_{ij}$,  (inclusive
of external  drives  where  applicable)  flowing  through  the  junction
between  sites $i$ and $j$, is viewed as consisting of three  `channels'
in  parallel:  superconductive,  resistive  ( or ohmic) and  capacitive.
The currents in each of these  channels can be expressed in terms of the
phase  difference,   $\theta_{ij}=   \theta_i   -\theta_j$,  across  the
junction.  This leads to the following equation for the evolution of the
latter in time:

\begin{equation}
\label{evolve0}
\frac{C \hbar}{2e}\frac{d^2\theta_{ij}}{dt^2}+
\frac{\hbar}{2eR}\frac{d\theta_{ij}}{dt}  +  I_c\sin\theta_{ij}  =i_{ij}
\end{equation}

Here $C$, $R$ and $I_c$  are the  shunt--capacitance,  shunt--resistance
and  critical  current of the junction  respectively.  We note that
Eq.(\ref{evolve0})  holds under  assumptions of zero  temperature,  zero
magnetic field and infinite  perpendicular  magnetic  penetration depth.
The last of these allows us to neglect self--induced magnetic fields.

Using total  current  conservation (TCC) at each site  \cite{Shenoy},  we can
write
\begin{equation}
\label{evolve}
\sum_{\langle ij \rangle}
\beta_c \frac{d^2\theta_{ij}}{d\tau^2} +
\frac{d\theta_{ij}}{d\tau} + \sin\theta_{ij} = I^{ext}_i \quad \forall i
\end{equation}
where $I^{ext}_i=  i^{ext}_i/I_c$ is the normalised current being fed to
or  extracted   from  the  array  site  $i$  (see  Fig.  ),  $\beta_c  =
2eI_cR^2C/\hbar$   is  the   McCumber--Stewart   parameter   and  $\tau=
t(2eI_cR)/\hbar$  is the time  measured  in units of the  characteristic
period $\omega_c^{-1}  =\hbar/(2eI_cR)$.  The summation on $j$, over the
nearest  neighbours of $i$, can alternatively be expressed in terms of a
multiplication by the discrete laplacian, $G_0^{-1}$.  Eq.[\ref{evolve}]
then assumes the matrix form
\begin{equation}
\label{under}
\sum_j  {\left (
G_0^{-1} \right)}_{ij} {\ddot{\theta}}_j =
\beta_c^{-1} [I^{ext}_i
-\sum_{\langle ij \rangle}
 {\dot{\theta}}_{ij} +
\sin\theta_{ij}] = -d_i
\end{equation}

If we set $\theta_i= x_i$ and  ${\dot{\theta}}_i=y_i$,  the complete set
of dynamical  equations reads ${\left ( {G_0}^{-1}  \right)}_{ij}  {\dot
{y}}_j = -d_i([x],[y])$ and ${\dot {x}}_i = y_i$.  For the overdamped  case,
corresponding to $\beta_c = 0$, the relevant equations are

\begin{equation}
\label{matrix}
	\sum_j {\left ( G_0^{-1} \right)}_{ij} { \dot{\theta}}_j = -d_i
\end{equation}
where  $-d_i$ is now  given by  $I^{ext}_i  -\sum_{\langle  ij  \rangle}
\sin\theta_{ij}$   but   $G_0^{-1}$  is,  of  course,  the  same  as  in
Eq.(\ref{under}).

It follows that for an $N_x \times N_y \equiv N$ array each  integration
time step of Eq.(\ref{matrix})  has a complexity ${\cal O} (N^2)$.  This
is   because  at  every   upgradation   of  the  $N$  state   variables,
${\dot{\theta}}$,  in the under--,  and  $\theta$,  in the  over--damped
case, the constant  $N\times N$ matrix $G_0$ has to be multiplied by the
divergence  vector $[d]$.  It was first noticed by Eikmans and Himbergen
\cite{Eikmans}  that the  form of the  $G_0^{-1}$  is such
that this  multiplication can actually be carried out in ${\cal O} (N\ln
N)$ steps.  The procedure  was  subsequently  improved upon by Dominguez
{\it et al.}  \cite{Doming} who combined the fast-fourier transformation
used by Eikmans  {\it et al.}  with the  method of cyclic  reduction  to
achieve a roughly 30\% increase in speed.

It is  noteworthy  that  these  algorithms  are  applicable  even in the
presence of an external  magnetic  field.  Indeed, the  application of a
field,  $B_0\hat{z}$,  perpendicular to the array  transforms the phases
$\theta_{ij}$ into the gauge-invariant  combinations  $\theta_{ij}+2 \pi
A_{ij}$  where  $A_{ij}  =1/\phi_0  \sum_i^j   \vec{A}\cdot   \vec{dl}$,
$\vec{A}$  is  the  vector  potential  and  $\phi_0=\hbar/(2e)$  is  the
elementary  flux  quantum  threading a  plaquette.  This  transformation
clearly  affects  only the  divergence  term, and leaves  untouched  the
matrix  $G_0^{-1}$, on whose form the algorithms  are based.  Similarly,
white noise, which is taken into account by  introducing a noise current
into  Eq.(\ref{under}), also modifies only the divergence and hence does
not  affect  the   applicability  of  these  algorithms.  The  statement
continues  to hold even if we grade  each  bond,  i.e.  make the $R$ and
$I_c$ junction--dependent \cite{Datta2}.

In this paper we extend  these fast  algorithms  firstly to arrays  with
busbars and secondly to those with defects in the form of missing bonds.
As has already been mentioned,  busbars often form the current injection
and/or   extraction   edges  of  experimental   arrays.  In  experiments
involving  vortices, which are repelled by busbars, these have been used
to produce collimated  vortex--streets  \cite{ZantEPL}.  The dynamics of
vortices have also been  investigated  with one edge shorted by a single
busbar  \cite{Ravi}.  Arrays with defects have  likewise  arisen in a
number  of  contexts.  The   breakdown   of   superconductive   flow  in
current--driven  arrays  with  linear  defects,  for  example,  has been
investigated  in some detail.  The  exploration of the  multiple--vortex
sector, which arises in this study, requires running on large arrays and
is  all  but   impossible   without  the  algorithm  we  discuss   below
\cite{Datta1}.  Defects can also be used to provide a collimated beam of
vortices \cite{Datta3}.

The paper is organised as follows.  In section \ref{sec:Eik}, we present
a  simplified  derivation  of  the   Eikmans--Himbergen   algorithm  and
interpret   all  the  key   equations  in  the   familiar   language  of
electrodynamics.  Apart  from being  more  transparent,  our  derivation
clarifies the role of boundaries and the connection  between the lattice
and continum  descriptions of the systems being studied.  These insights
are used in section  \ref{sec:busbar}  to generalize  this  algorithm to
arrays with busbars.  The case of single  busbar  forces us to resort to
the technique of cyclic reduction, which we consequently discuss at this
point.  In section  \ref{sec:defect}, we extend these algorithms to JJAs
with  defects,   created  by   eliminating   or  adding  bonds.  Section
\ref{sec:conclu} contains a summary and discussion of our results.

\section{The Eikmans--Himbergen algorithm}
\label{sec:Eik}
\subsection{Preliminaries}
\label{subsec:prelim}

  All   efficient   algorithms   for  the   numerical   integration   of
  Eq.(\ref{matrix}) make essential use of the properties of the discrete
  laplacian,  $G_0^{-1}$.  We thus begin by listing  the more  important
  properties  of this  matrix.  From its  definition  (as  given by Eqs.
  (\ref{under} and \ref{matrix}), it follows that $G_0^{-1}$  specifies
  the connectivity  between different sites.  More precisely,  ${\left (
  G_0^{-1}\right  )}_{ij}=  {\left(  G_0^{-1}\right  )}_{ji}=  -1 $  for
  $i\neq  j$ (site  $i$  connected  to site  $j$) and 0  otherwise.  The
  diagonal  element ${\left  (G_0^{-1}\right  )}_{ii}=$  total number of
  points to which the site $i$ is connected, and hence $Tr(G_0^{-1})$ is
  twice the number of bonds in the network.

  Clearly,   $G_0^{-1}$  is  real  and   symmetric.  As  a  result,  its
  eigenvalues   are   real   and   its   eigenbasis   complete   in  the
  $N$--dimensional space of states.  Furthermore,  $\det(G_0^{-1}) = 0$.
  This  can  be   deduced  as   follows.  We  first   note  that   since
  Eq.(\ref{evolve}) involves only the phase {\it differences} (and their
  time   derivatives)  it  is  invariant  with  respect  to  the  global
  transformation    ${\dot{\theta}}_i    \rightarrow    {\dot{\theta}}_i
  +\alpha(\tau)    \quad    \forall    i$.   Substituting    this   into
  Eq.(\ref{matrix}), we have

  \[
  \sum_j {\left( G_0^{-1}\right )}_{ij}  ({\dot{\theta}}_j+\dot{\alpha}(
  \tau))= -d_i = \sum_j {\left( G_0^{-1}\right  )}_{ij}  {\dot{\theta}}_j
  \]

   This implies that  $\dot{\alpha}(\tau ) \sum_j {\left( G_0^{-1}\right
   )}_{ij}  =0 \quad  \forall  i$.  I.e., if we fix $i$ and sum over all
   the  columns  $j$, we get zero.  (The same is true, of  course, if we
   fix $j$ and sum over $i$ since the matrix is symmetric).  Hence $\det
   (G_0^{-1})   =0$  and  only   $(N-1)$   of  the  $N$   equations   in
   Eq.(\ref{evolve})  are independent.  (We could,  alternatively,  note
   that  $(\Psi_0)_i = 1 \forall i$ is an eigenvector of $G_0^{-1}$ with
   eigenvalue 0).

   To explicitly evaluate  ${\dot{\theta}}_i$ from $\theta_i$ we have to
   eliminate the extra degree of freedom.  For what immediately follows,
   the  most  convenient   choice  is  $\sum_i   {\dot{\theta}}_i   =0$.
   Eq.(\ref{matrix}) is then replaced by
   \begin{equation}
   \sum_j {\left( {\cal G}_0^{-1} \right )}_{ij} {\dot{\theta}}_j = -D_i
   \end{equation}
   where  $D_i=d_i,  \; i=1  \cdots  (N-1)$ and $D_N =0$ while  ${\left(
   {\cal G}_0^{-1} \right)}_{ij}= {\left( {G}_0^{-1} \right )}_{ij}\quad
   i=1\cdots   (N-1),\;   j=1\cdots  N$  and  ${\left(  {\cal  G}_0^{-1}
   \right)}_{Nj}=1  \forall j$.  Moreover, the vector $[d]$ must satisfy
   $\sum_i d_i =0$ as can be seen by performing an  additional  sum over
   $i$  in   Eq.(\ref{matrix})   and   using  the  fact   that   $\sum_i
   {(G_0^{-1})}_{ij}=0$.  For  JJAs  this  condition  is   automatically
   satisfied  since $\dot  \theta_{ij}$  and  $\sin\theta_{ij}$  are odd
   functions of their arguments and the net external  current fed to the
   array is zero.

   One can, in principle, invert ${\cal G}_0^{-1}$ and determine $\left[
   \dot{\theta} \right]$ from $[D]$.  This is the usual ${\cal O} (N^2)$
   process.  We shall refrain from adopting this  procedure  and return,
   instead, to working with Eq.(\ref{evolve}).

   In  doing  so,  we  shall  find  it  useful  to  keep  in  mind  some
   electrostatic  analogs of the equations we happen to be dealing with.
   These emerge clearly if we write the current conservation equation at
   the site $k$ as

	\begin{equation}
	I_{k,\mu}^{total}= I_{k,\mu}^{s}+ I_{k,\mu}^{n}+ I_{k,\mu}^{ext}
	\end{equation}

    Here $\mu=1,2$  represents the $x$ and $y$ directions  respectively.
    Furthermore,           $I_{k,\mu}^{total}$,          $I_{k,\mu}^{s}=
    \sin\Delta_\mu\theta_k$, $I_{k,\mu}^{n}=\Delta_\mu {\dot{\theta}}_i$
    and $ I_{k,\mu}^{ext}$  are the total,  superconducting,  normal and
    external  currents,  respectively,  at  node  $k$ in the  $\mu^{th}$
    direction.  Using the TCC condition, $\nabla\cdot I_i^{total}=0$, we
    have
	\begin{equation}
	\label{laplace}
	\nabla^2 {\dot{\theta}}_i = \nabla\cdot \left (
	{\vec{I}}_i^s+{\vec{I}}_i^{ext} \right )
	\end{equation}

    On comparing Eq.(\ref{laplace}) with the Laplace Equation $\nabla^2
    \phi    =     -\rho_{total}/\epsilon_0$     and    recalling    that
    $\rho_{total}=\rho_{free}+\rho_{bound}$,  we arrive at the following
    correspondences
	\begin{eqnarray}
	\vec{D}(\vecr ) & \equiv & {\vec{I}}_i^{ext} \\
	- \vec{P}(\vecr ) & \equiv & {\vec{I}}_i^{s} \\
	\phi (\vecr ) & \equiv & {\dot{\theta}}_i
	\end{eqnarray}

    Thus  the  condition  imposed  above  on the  ${\dot{\theta}}_i$  is
    nothing  other than a choice of reference  potential  or gauge.  We,
    furthermore,  note that the JJA thought of as a dielectric medium is
    highly non--linear.  Indeed,  $P_{k,\mu}\equiv  \sin(-\int E_{k,\mu}
    dt)$,  where   $E_{k,\mu}  =   \Delta_\mu{\dot{\theta}}_k$   is  the
    $\mu^{th}$  component of the electric field at site $k$.  This is to
    be contrasted with $ \vec{P}(\vecr ) \propto  \vec{E}(\vecr )$ for a
    linear dielectric.

   \subsection{ The ${\cal O}(N\ln N )$ procedure}
	\label{subsec:nlogn}

    Going back to  Eq.(\ref{under}),  $G_0^{-1}$  can  alternatively  be
    inverted by defining the Green's function [see Jackson]
    \begin{equation}
    \label{tgreen}
    \tilde{G} (\vecr,\vecrp ) = \sum_{\lambda_\veck \neq 0}
     \frac{1}{\lambda_\veck } \Psi_\veck (\vecr)\Psi_\veck^\dagger (\vecrp)
    \end{equation}
   where  $\lambda_\veck$ are the eigenvalues and $ \Psi_\veck  (\vecr)$
   the orthonormalised eigenvectors of $G_0^{-1}$, i.  e.
	\begin{equation}
	\label{eigen}
	G_0^{-1} \Psi_\veck (\vecr)=  \lambda_\veck  \Psi_\veck (\vecr)
	\end{equation}
    and $\vecr$ are points on the  lattice.  More  explicitly,  the site
    coordinates $(x_i,\; y_i) \equiv {\vec{r}}_i$  referred to an origin
    located at the lower left hand corner of the lattice, are related to
    the site  indices  $i$ used  earlier  by $i = x_i + y_i N_x +1$.  We
    note that to get a well-defined  function, we must explicitly remove
    from the sum in Eq.(\ref{tgreen}),  the zero mode, which necessarily
    exists, since $G_0^{-1}$ is singular.

    The  removal  of this  mode,  however,  creates  a  slight  problem.
    Indeed, if we premultiply by $G_0^{-1}$ and use Eq.(\ref{eigen}) we
    get
	\begin{eqnarray}
	\sum_{\vecrrp } G_0^{-1} (\vecr ,\vecrrp )
	\tilde{G} (\vecrrp, \vecr )& = &
	  \sum_{\veck \neq 0}\frac{1}{\lambda_\veck}\lambda_\veck
	\Psi_\veck (\vecr ) \Psi^\dagger_\veck (\vecrp )  \\ \nonumber
	  &= & \sum_{\veck } \Psi_\veck (\vecr )
	  \Psi^\dagger_\veck (\vecrp ) -
	  \Psi_0 (\vecr ) \Psi^\dagger_0 (\vecrp )  \\ \nonumber
	& =& \delta (\vecr - \vecrp ) -\frac{1}{\sqrt{L_xL_y}} \left (
	\begin{array}{c} 1 \\ \vdots \\ 1 \end{array} \right )
	\frac{1}{\sqrt{L_xL_y}} \left ( 1, \; 1 \cdots 1 \right )
	\end{eqnarray}
	which in discretised form is
	\begin{equation}
        \label{criteriong}
		  \sum_j  {\tilde{G}}_{ij}  {\left( {G}_0^{-1}
        \right)}_{jk}= \delta_{ik} - \frac{1}{L_x L_y}
	\end{equation}

    Now if we multiply  Eq.(\ref{matrix})  from the left by $\tilde{G}$
    we get ${\dot{\theta}}_i - \frac{\sum_i {\dot{\theta}}_i} {L_x L_y}$
    which reads just right, viz.
	\begin{equation}
	\label{matrixf}
	{\dot{\theta}}_i = {\tilde{G}}_{ij} d_j
	\end{equation}
    {\it   provided}   the    ${\dot{\theta}}_i    $   satisfy   $\sum_i
    {\dot{\theta}}_i  =0$.  In other  words, with an  appropriate  gauge
    choice,  ${\tilde{G}}$ does allow us to invert  $G_0^{-1}$.  Indeed,
    the requirement  that it does so fixes the gauge uniquely.  It is to
    be noted that Eq.(\ref{matrixf}) involves $[d]$ and not $[D]$.

    The function,  $\tilde{G}(\vecr  ,\vecrp )$, can be easily evaluated
    for a periodic lattice.  The form of the operator  $G_0^{-1}(\vecr ,
    \vecrp )$, in this case, is given by
	\begin{equation}
	\label{pergreen}
	G_0^{-1}(\vecr , \vecrp )=
	4\delta_{\vecr, \vecrp } - \delta_{\vecr\pm {\hat{e}}_x,\vecrp }
	- \delta_{\vecr\pm {\hat{e}}_y,\vecrp }
	\end{equation}
    It is easily  checked that the  normalised  eigenvectors  are of the
    form $\frac{\exp{i\veck  \cdot\vecr}}{\sqrt{L_xL_y}}$ and correspond
    to the eigenvalues
	\begin{equation}
	\label{slambda}
	\lambda_\veck = 4-2\cos k_x-2\cos k_y
	\end{equation}
    We note that $\lambda_{\veck = 0}=0$.

    Due to the rectangular  periodicity of the lattice the  eigenvectors
    must  satisfy   $\Psi_\veck   (\vecr  )=   \Psi_\veck   (\vecr  +L_i
    {\hat{e}}_i   )  \quad  i=x,y$.  To  fulfil  this   condition,   the
    wavevectors  must in turn be  quantised  as  $k_i=(2n\pi)/L_i  \quad
    i=x,y$  and  $n=0,1,\cdots   (L_{i-1})$.  The  Green's  function  of
    interest is consequently
	\begin{equation}
	\tilde{G}(\vecr ,\vecrp ) =
	\frac{1}{L_xL_y}
	\sum_{\veck \neq 0}
	\frac{1}{4-2\cos k_x -2\cos k_y}
	\exp{i\veck\cdot (\vecr -\vecrp )}
	\end{equation}

    We can now  estimate  the  number  of  steps  required  to  evaluate
    $\sum_{\vecrp}  \tilde{G}(\vecr  ,\vecrp ) d(\vecrp)$.  The sum over
    $\vecrp$ is a fourier  transform  and can be carried out in $ N lnN$
    steps  to  produce  $d(\veck)$.  Then  come the $N$  multiplications
    $\lambda_\veck d(\veck)$ followed by the sum over $\veck$, which has
    the form of an inverse fourier transform, and requires an additional
    $N ln N$ steps, making a grand total of $N(2lnN+1)$ multiplications.

    Quite generally then, ${\dot{\theta}}(\vecr )$ can be evaluated from
    Eq.(\ref{matrixf})  by (i)  creating  an  $L_x  \times  L_y$  matrix
    $G_0(\veck  ) =  N_\veck  /\lambda_\veck$  where  $N_\veck$  are the
    normalisation  factors due to $\Psi_\veck (\vecr )$, (ii) evaluating
    $d(\veck )=  \sum_{\vecrp  }d(\vecrp )  \Psi_\veck  (\vecrp )$, this
    being the forward  transform  (W), (iii)  computing  $d^\prime (k) =
    G_0(\veck  ) d(\veck )$, and (iv) taking the 2-D  inverse  transform
    $\tilde{W}$  to  get  $\dot{\theta}(\vecr  ) =  \sum_\veck  d^\prime
    (\veck ) \Psi_\veck  (\vecr )$.  For certain types of transforms and
    certain values of $L_x$ and $L_y$, $W$ and  $\tilde{W}$  can each be
    performed by matrix  decomposition or row--column  techniques with a
    complexity ${\cal O} (N\ln N)$.  We now turn to a discussion of this
    procedure in the context of finite arrays.

    \subsection{The Imposition of Boundaries}
	\label{subsec:bound}

    For   a   finite    array,   the   general   form   of    $G_0^{-1}$
    (Eq.(\ref{pergreen}))  is  modified  since now fewer than four bonds
    meet at all sites on the boundary  (see  Fig.(1)).  For  non--corner
    points on the left edge of the array for example, the  operator  has
    the    form    $3\delta_{\vecr,    \vecrp    }    -    \delta_{\vecr
    +{\hat{e}}_x,\vecrp  }  -  \delta_{\vecr\pm  {\hat{e}}_y,\vecrp  }$.
    However, if $\Psi$ satisfies $\Psi  (x_0,y)-\Psi  (x_0-1,y) =0 \quad
    \forall y$ along this edge, one can continue using the periodic form
    of   $G_0^{-1}$   because   then   $(4\delta_{\vecr,   \vecrp   }  -
    \delta_{\vecr\pm    {\hat{e}}_x,\vecrp    }    -    \delta_{\vecr\pm
    {\hat{e}}_y,\vecrp  })  \Psi(x_0,y)$  automatically  reduces  to  $(
    3\delta_{\vecr,  \vecrp } -  \delta_{\vecr  +{\hat{e}}_x,\vecrp  } -
    \delta_{\vecr\pm  {\hat{e}}_y,\vecrp  })  \Psi(x_0,y)$.  We thus see
    that the  finiteness  of the array  imposes the  following  boundary
    conditions on $\Psi$
    \begin{mathletters}
    \label{bc}
    \begin{equation}
    \label{bca}
    \Psi (x_0,y) -\Psi  (x_0-1,y) =0 \quad  \forall y
    \end{equation}
    \begin{equation}
    \label{bcb}
    \Psi (x,y_0) -\Psi  (x,y_0-1) =0 \quad \forall x
    \end{equation}
    \begin{equation}
    \label{bcc}
    \Psi (x_L,y) -\Psi  (x_L+1,y) =0 \quad \forall  y  \quad
	\mbox{where}\quad   x_L=x_0+L_x  -1
    \end{equation}
    \begin{equation}
    \label{bcd}
    \Psi (x,y_L) -\Psi  (x,y_L+1) =0 \quad \forall  x  \quad
    \mbox{where} \quad  y_L=y_0+L_y  -1
    \end{equation}
    \end{mathletters}
     where  $(x_0,y_0)$  and  $(x_L,y_L)$  are the  diagonally  opposite
     corners of the rectangular lattice.  It is amusing to note that the
     left--hand  side of  Eqs.(\ref{bca}-  \ref{bcd})  are the  discrete
     derivatives of $\Psi(x,y)$ along the edges  $x=x_0$,$y=y_0$,$x=x_L$
     and  $y=y_L$,  respectively.  Thus the finite  case  poses for us a
     discretised Neumann boundary value problem.

     Furthermore, we note that $\lambda_\veck$  (see  Eq.(\ref{slambda}))
     has  the   symmetries   of  the  square,  i.e.  $\lambda_{k_x,k_y}=
     \lambda_{\pm  k_x,\mp k_y}=  \lambda_{-k_x,-k_y}$.  Thus any linear
     combination of the  corresponding eigenvectors,
     \begin{eqnarray}
     \Psi  (x,y)&=&  a_1   \exp{i(k_xx+k_yy)}+a_2 \exp{i(-k_xx+k_yy)}\\
	 			\nonumber
		& &+ a_3 \exp{i(k_xx-k_yy)}+a_4 \exp{i(-k_xx-k_yy)}
      \end{eqnarray}
     is  an  eigenvector   of  $G_0^{-1}$   with  the  same   eigenvalue
     $\lambda_\veck$. To find the specific linear
     combination,  which satisfies the boundary  conditions given above,
     we need merely impose
     each  of  Eqs.(\ref{bca}-\ref{bcd})  in  turn.
    Eq.(\ref{bca}) can be satisfied by choosing
    %
    %
   $x_0=1/2$,  $a_1=a_2$, $a_3=a_4$
    and the resulting wavefunction is
    \begin{equation}
    \label{aa}
    \Psi (x,y)=  2a_1\cos  k_xx \exp  (ik_yy)+
		 2a_3\cos  k_xx \exp  (-ik_yy)
    \end{equation}
    Subjecting  Eq.(\ref{aa})
    next to the constraint Eq.(\ref{bcb}), we get
    \begin{equation}
    \Psi (x,y)= 4a_1\cos k_xx \cos k_yy
    \end{equation}
    where $y_0$ is now required to be $1/2$ as well. The fact that $x_0 =
	y_0= 1/2$ means that the
	origin of coordinates is chosen on the dual lattice and both
	coordinates of every lattice point are half integers.

    The boundary  conditions  Eq.(\ref{bcc})  and  Eq.(\ref{bcd}) at the
    right and upper edges of the array can be  satisfied  by  imposing a
    quantisation condition on $\veck$.  Indeed, using Eq.(\ref{bcc}) we
    have
      \begin{equation}
      -8a_1\cos k_yy \sin(k_xL_x)\sin(\frac{k_x}{2}) =0 \quad \forall y
      \end{equation}
    whereby $k_x= n_x\pi/L_x, \quad n_x=0,1,\cdots  ,L_x-1$.  Similarly
    using Eq.(\ref{bcd}) $k_y= n_y\pi/L_y, \quad n_y=0,1,\cdots ,L_y-1$.

    Thus finally, the  orthonormalised  wavefunctions we are looking for
    are
    \begin{equation}
    \Psi (x,y) =
    \sqrt{\frac{2}{L_xL_y}} \sqrt{\left (1-\frac{1}{2}\delta_{k_x,0}
    -\frac{1}{2}\delta_{k_y,0}\right )}  \cos(k_xx) \cos (k_yy) \quad
    \veck \neq  0
     \end{equation}

     The resulting Green's function to be used in  Eq.(\ref{tgreen})  is
     thus \cite{Eikmans}
     \begin{equation}
     \tilde{G}(\vecr  ,\vecrp  )
     \frac{4}{L_xL_y}
     \sum_{\veck \neq 0}
     \frac{(1-\frac{1}{2}\delta_{k_x,0}
     -\frac{1}{2}\delta_{k_y,0})}  {4-2\cos  k_x -2\cos k_y}
     \cos(k_xx)   \cos   (k_yy)    \cos(k_xx^\prime)
     \cos    (k_yy^\prime)
     \end{equation}

	 The action of this Green's function on a uniform current drive,
	 $I^{ext}(x,y) = I^{ext} (\delta_{x,x_0} - \delta_{x,x_{L_x}})$ can be
	 rigorously shown to be \cite{thesis}
	 \begin{equation}
	 \label{current1}
	 \sum_{y'} \tilde{G}(x,y,x',y') I (x',y')= I (\frac{L_x}{2} -x)
	 \end{equation}

     \section{Inclusion of Bus--bars}
     \label{sec:busbar}

    Using the  techniques  of the  previous  section we  generalise  the
    procedure  to  arrays  with  either  a  single  or a  double  busbar
    (henceforth referred to as SBB and DBB respectively).

    In  the  SBB  case,  the  busbar  is  normally   placed   along  the
    current--extraction--edge (see Fig.(2)).  The corresponding $\theta$
    and  $\dot{\theta}$  are zero for all time i.e.  they do not evolve.
    The drive edge is  however  open and can be  connected  to {\it any}
    kind of current  profile.  Such an array, with a linear profile, has
    been used previously to study vortex  dynamics  \cite{Ravi}.  On the
    other hand, an DBB array has shorts  along a pair of parallel  edges
    (say those at $x=x_0$ and $x=x_L+1$ as in Fig(3)).  In this case, no
    current  can be  injected  in the  $x$--  direction  but  the  $y$--
    direction is available to an arbitrary current drive.  Such an array
    with electrically  connected  busbars could be used to simulate e.g.
    the ``channel'' in the experiments conducted by Van-der-Zant {\it et
    al.}\cite{ZantEPL}.

    We note that by setting the $\dot{\theta}$  along the busbar to zero
    and  measuring  all other  $\dot{\theta}$'s  with  respect  to these
    preassigned  variables,  we  are  unambigiously  fixing  the  gauge.
    Furthermore,  since the  wavefunction  in a  rectangular  system  is
    separable, i.e.  $\Psi (x,y) = \Psi_1 (x)  \Psi_2(y)$, any change in
    the boundary  conditions along the $x$-- direction, has no impact on
    $\Psi_2(y)$,  which therefore  continues to have the form derived in
    Setion II, viz.
    \[ \Psi_2(y)= \sqrt{\frac{2}{L_y}}
    \sqrt{\left (1-\frac{1}{2}\delta_{k_y,0}\right )} \cos k_yy \]

    The altered b.c.s however enter crucially into the  determination of
    $\Psi_1(x)$ and hence in to that of $\Psi (x,y)$.

	\subsection{Arrays with a Single Bus-bar}
	\label{subsec:sbb}
	For the SBB case,  we have the following b.c.s
	\begin{mathletters}
	\label{bcSBB}
	\begin{equation}
	\Psi_1 (x_0) -\Psi_1 (x_0-1) =0 \quad \forall y
	\label{bcSBBa}
	\end{equation}
	\begin{equation}
	\Psi_1 (x_L+1) =0  \quad \forall y \quad
	\mbox{where}\quad   x_L=x_0+L_x  -1
	\label{bcSBBb}
	\end{equation}
	\end{mathletters}

    The second of these results from the fact that the wavefunction must
    vanish  at  the   shorted   edge  as   discussed   above.  Condition
    Eq.(\ref{bcSBBa})  fixes the form of  $\Psi_1(x)$  to be $\sim  \cos
    k_xx$  where $x$ is a  half--odd  integer,  while  Eq.(\ref{bcSBBb})
    quantises  $k_x$:  $k_x=((2n_x+1)\pi)/(2L_x+1)\;  \;n_x = 0,1,\cdots
    ,L_x-1$.  The  normalisation  constant  is fixed by the  requirement
    that $A^2\sum_{p=1}^{L_x} \cos^2 k_x(p-\frac{1}{2}) =1 $.  Thus
	\begin{equation}
	\Psi_1 (x)= \sqrt{\left (\frac{2}{L_x + \frac{1}{2} }\right )}
	\cos k_xx
	\end{equation}
   and hence the Green's  function  $\tilde{G}$ for the case of a single
   busbar is
	\begin{equation}
	\tilde{G}(\vecr ,\vecrp ) =
	\frac{4}{L_y(L_x+\frac{1}{2})}
	\sum_{\veck}
	\frac{(1-\frac{1}{2}\delta_{k_y,0})}{4-2\cos k_x -2\cos k_y}
	\cos(k_xx) \cos (k_yy) \cos(k_xx^\prime) \cos (k_yy^\prime)
	\end{equation}
	This $\tilde{G}$ coincides with the $G_0$ for this case because the
	zero--mode is absent.

   The action of  $\tilde{G}$ on a uniform input drive  $I^{ext}(x,y)  =
   I^{ext}  \delta_{x,x_0}$ can, as in the case of  Eq.(\ref{current1}),
   be exactly shown to be \cite{thesis}
   \begin{equation}
   \sum_{y'} \tilde{G}(x,y,x',y') I (x',y')= I (L_x - x + \frac{1}{2})
   \end{equation}

   Although a 1D Fast Cosine transform can be used along the $y$-- axis,
   this  cannot  be done  for the  $x$--direction,  since no {\it  fast}
   Cosine transform of the form
   \[  C(m)=  \sum_{n=0}^{L-1}
        	\tilde{C}(n)\cos  \frac{(2n+1)(2m+1)\pi}{(4L+2)}  \]
   is  known to  exist  \cite{Cosine}.  If we  resort  to  usual  matrix
   multiplication along ${\hat{e}}_x$, the algorithm becomes ${\cal O}(2
   N N_x+2N  \ln  N_y+N)$ in  complexity  and is useful  only if $N_y >>
   N_x$.

    A  much  more  efficient  approach  is  to  combine  the  FCT  along
    ${\hat{e}}_y$  with cylic reduction along the  $x$--direction.  This
    can  be  carried  out as  follows  \cite{Recipes}.  We  begin  by
    writing Eq.(\ref{matrix}) as
	\begin{equation}
	\dot{\theta}_{x-1,y} + \dot{\theta}_{x,y^\prime}  T_{y^\prime,y} +
	\dot{\theta}_{x+1,y} = -d_{x,y}
	\end{equation}
    where the operator $T_{y^\prime,y}$ is defined to be $T_{y^\prime,y}
    =     \delta_{y^\prime,y-1}     +      \delta_{y^\prime,y+1}      -4
    \delta_{y^\prime,y}$
    Taking the Cosine  transform  along
    the  $y$--   direction  i.\  e.\  setting   $\dot{\theta}_{x,y}   =
    \sum_{k_y} \dot{\theta}_{x,k_y} \cos k_y y$, we find that
	\begin{equation}
	 \dot{\theta}_{x,y^\prime}  T_{y^\prime,y} = \sum_{k_y} 2 (\cos k_y -2)
	 \dot{\theta}_{x,k_y} \cos k_y y
	\end{equation}

   As    a    consequence     the    $y$--    cosine     transform    of
   $\dot{\theta}_{x,y}$ satisfies
	\begin{equation}
	\label{eq:dotky1}
	\dot{\theta}_{x-1,k_y} + \lambda {(k_y)} \dot{\theta}_{x,k_y}  +
	\dot{\theta}_{x+1,k_y} = -d_{x,k_y}
	\end{equation}
   where  we have  used  $\lambda({k_y})  =  2(\cos  k_y -2 )$.  Writing
   Eq.(\ref{eq:dotky1})  thrice with $x$ set equal to  $x-1$,  $x$
   and $x  +1$  respectively,  multiplying  the  second of these equations  by
   $-\lambda(k_y)$ and adding all the three of them together, we get
	\begin{equation}
	\label{eq:dotky2}
	\dot{\theta}_{x-2,k_y} + \lambda^{(1)} {(k_y)} \dot{\theta}_{x,k_y}  +
	\dot{\theta}_{x+2,k_y} = -d_{x-1,k_y}+\lambda (k_y) d_{x,k_y}
	-d_{x+1,k_y} = -d_{x,k_y}^{(1)}
	\end{equation}
    where $ \lambda^{(1)}(k_y) = 2- {\left (  \lambda(k_y)\right  )}^2$.
    We note  that  the  $x$  values  occurring  in  Eq.(\ref{eq:dotky2})
    increase    in    steps    of   $2$   as    opposed    to   $1$   in
    Eq.(\ref{eq:dotky1}). The  two  equations  are,  however, of the same
    form.  We can thus use the  reduction  procedure  repeatedly  to get
    equations with $x$--values increasing in steps of $4,8,\cdots$ until
    after $m$ stages ($2^{m}=L_x$), we end up with a single equation for
    the central line of variables:
	\begin{equation}
	\label{eq:central}
	\lambda^{(m-1)} (k_y) \dot{\theta}_{x_0+L_x/2,k_y} =
d^{(m-1)}_{x_0+L_x/2,k_y}-
	\dot{\theta}_{x_0,k_y} -\dot{\theta}_{x_L+1,k_y}
	\end{equation}

    The  $\lambda^{(p)}(k_y)$  and $d^{(p)}_{x,k_y}$  occurring in these
    equation are defined by the recursive relations:
    \begin{equation}
    \lambda^{(p)}(k_y) = 2- {\left (\lambda^{(p-1)}(k_y) \right )}^2
    \end{equation}
    \begin{equation}
    d^{(p)}_{x,k_y} =  d^{(p-1)}_{x-2^{p-1},k_y} - \lambda^{(p-1)}(k_y)
    d^{(p-1)}_{x,k_y} + d^{(p-1)}_{x+2^{p-1},k_y}
    \end{equation}

    Thus, if $\dot{\theta}_{x_0,k_y}$ and  $\dot{\theta}_{x_L+1,k_y}$  are
    known, then  $\dot{\theta}_{x_0+L_x/2,k_y}$  can be immediately deduced.
    The  knowledge  of  $\dot{\theta}_{x_0+L_x/2,k_y}$   leads  to  that  of
    $\dot{\theta}_{x_0+L_x/4,k_y}$ and $\dot{\theta}_{x_0+3L_x/4,k_y}$  which
in
    turn determine  $\dot{\theta}_{x_0+qL_x/8,k_y}$ ($q = 1,3,5,7$) etc.  As
    for the starting values,  $\dot{\theta}_{x_L+1,k_y}=0  \; \forall \;
    k_y$, since this is the y--cosine transform of  $\dot{\theta}_{x,y}$
    for $x=x_L+1$, i.e.  taken along the busbar.  $\dot{\theta}_{x_0,k_y}$
    must however be determined by explicit multiplication.  Fortunately,
    since we know the relevant Green's function explicitly, we can carry
    this out in ${\cal O}(N)$  steps, by taking the cosine  transform of
    Eq.(\ref{matrixf}):
    \begin{equation}
    \dot{\theta}_{x_0,k_y}=
    \frac{4}{L_y(L_x+\frac{1}{2})}
	\sum_{k_x} \sum_{x^\prime}
	\frac{\left (1-\frac{1}{2}\delta_{k_y,0}\right )}
	{4-2\cos k_x -2\cos k_y }
	\cos k_x x_0 \cos k_x x^\prime d_{x^\prime,k_y}
    \end{equation}

    The  number of steps  required  to  evaluate  all the  higher--order
    divergences  $d^{(p)}_{x,k_y}, \; p=1,\cdots m-1$ used by the method
    can be shown to be  $N_y\left  \{  (N_x/2)  - \ln_2 N_x  \right  \}$
    \cite{thesis}   .  With  these   divergences  in  hand,  the  cyclic
    determination of  ${\dot{\theta}}_{x,k_y}$  has a complexity  ${\cal
    O}(N)$.  Adding to all this, the number of multiplications  involved
    in carrying out the forward and backward  FCTs in the  y--direction,
    we conclude that the entire  procedure  can be carried out in ${\cal
    O} ( 2N \ln N_y +2N +N/2 - N_y \ln_2 N_x)$ steps.

    \subsection{The Case of a Double Bus-bar}
	\label{subsec:dbb}
	Turning now to the DBB case, we note that the wave-function,
	$\Psi_1 (x)$, must
	now satisfy
	the following b.c.s
	\begin{mathletters}
	\label{bcDBB}
	\begin{equation}
	\Psi_1 (x_0) =0 \quad \forall y
	\label{bcDBBa}
	\end{equation}
	\begin{equation}
	\Psi_1 (x_L+1) =0  \quad \forall y
	\label{bcDBBb}
	\end{equation}
	\end{mathletters}

    Condition  Eq.(\ref{bcDBBa})  fixes  the form of  $\Psi_1(x)$  to be
    $\sim  \sin  k_xx$   where  $x$  is  an  integer   while   condition
    Eq.(\ref{bcDBBb})  restricts $k_x$ to the values  $(n_x\pi)/(L_x+1),
    \; \; n_x = 1,\cdots L_x$.  The normalisation  constant, A, is fixed
    by    $A^2\sum_{p=1}^{L_x}\sin^2    k_xp    =1$.    The    resulting
    orthonormalised wavefunction is then
	\begin{equation}
	\Psi_1(x)=  \sqrt{\frac{2}{L_x+1}} \sin k_xx
	\end{equation}
	and hence $\tilde{G}$ for an array with a double bus--bar is
	\begin{eqnarray}
	\tilde{G}(\vecr ,\vecrp ) & = &
	\frac{4}{L_y(L_x+1)}
	\sum_{\veck}
	\frac{(1-\frac{1}{2}\delta_{k_y,0})}{4-2\cos k_x -2\cos k_y}
	\sin(k_xx) \cos (k_yy)
	\sin(k_xx^\prime) \cos (k_yy^\prime)
	\end{eqnarray}
    In this case, the transformations along both directions are amenable
    to fast, i.e.  ${\cal O}(N\ln N)$, procedures  provided  $L_x+1$ and
    $L_y$ are both integral powers of 2.

    \section{Defects}
    \label{sec:defect}

     We move on next  onto  arrays  containing  defects  in the  form of
     broken  bonds.  As  mentioned  in Sec.  \ref{sec:intro},  arrays of
     this sort have recently  attracted the attention of several groups.
     Xia and Leath  showed that in  contrast  to networks  of  resistors
     \cite{Duxbury},  where  breakdown  emanates  from the
     most  critical  defect,  JJAs  with  missing  bonds  do not  become
     resistive, the moment any one junction  turns  critical.  Cohn {\it
     et  al.}  \cite{Cohn}  showed  that  defects  exert  an  additional
     pinning force on a vortex placed in the system.  Datta {\it et al.}
     \cite{Datta1} demonstrated the existence of multiple vortex sectors
     in the steady state  configurations  of the system.  They did so by
     running  on large  arrays  using a  modified  version  of the above
     algorithm which we now describe in detail.

     The  evolution  equation  $\dot{\theta}$  for an array  containing a
     single  missing  bond  between the sites $k \equiv  (x_0,y_0)$  and
     $(k+1) \equiv  (x_0+1,y_0)$ can be written in matrix notation ( see
     Eq.(\ref{evolve})) as
     \begin{equation}
     \label{evolve1}
     (G_0^{-1} +h) [\dot{\theta}] = [d]
     \end{equation}
     where $G_0^{-1}$ is the discrete Laplacian with free, periodic,
     SBB or DBB boundary conditions and $h$ is given by
     \begin{eqnarray}
     \label{hform}
     h_{k,k} = & h_{k+1,k+1} = & 1 \\ \nonumber
     h_{k,k+1} = & h_{k+1,k} = & -1 \\ \nonumber
     h_{i,j} = & 0 & i,j \neq k, k+1
     \end{eqnarray}
     Multiplying Eq.(\ref{evolve1}) with $\tilde{G}$ from the left we get
     \begin{equation}
     (I+A) [\dot{\theta}] = \tilde{G}[d]=\xi
     \end{equation}
     where  $A=\tilde{G} h$ and $I$ is the identity matrix.  The process
     of determining  $\tilde{G}  [d] =\xi$, as already been outlined and
     can be  executed  in $N\ln N$  steps.  One then  needs to  evaluate
     ${(1+A)}^{-1}  \xi$ efficiently.  To this end, we note that $A$ has
     the form
     \begin{equation}
     \label{Aform}
     A_{ij}=x_i ( \delta_{j,k} - \delta_{j,k+1} )
     \end{equation}
     where the numbers $x_i$ are got by explicitly by multiplying  $G_0$
     and $h$.  From Eq.(\ref{Aform}) it follows that
     \begin{equation}
     \label{Arec}
     A^q= A {(\Delta e)}^{q-1}
     \end{equation}
     where $ \Delta e = x_k -  x_{k+1}  $.  In fact,  $\Delta  e$ is the
     only  non-zero  eigen  value of the  matrix  $A$  and for the case
     discussed in Section \ref{subsec:bound} it is given by
     \begin{equation}
     \Delta e =
	\frac{4}{L_xL_y}
	\sum_{\veck \neq 0}
	\frac{\left ( 1-\frac{1}{2}\delta_{k_x,0} -\frac{1}{2}\delta_{k_y,0}
	\right )}  {4-2\cos  k_x -2\cos k_y} \\  \nonumber
	{(\cos(k_x(x_0+1)) -  \cos(k_xx_0))}^2 \cos^2   (k_yy_0)
     \end{equation}
     where $k_i=(n_i\pi )/L_i, \; n_i = 0 \ldots L_i-1$ and $i=x, y$.
     From Eq.(\ref{Arec}) it follows that
     \begin{equation}
     \label{taylor}
     {(I+A)}^{-1} = I - A + A^2 - A^3 + \cdots  = I - \frac{A}{1+\Delta e }
     \end{equation}

     Since the  constant  matrix $R= - A/(1 + \Delta e )$ contains  only
     two non--zero  columns, which are moreover negatives of each other,
     $[\dot{\theta}  ]$  can  be  determined  from  $\xi$  in  just  $N$
     multiplicative steps.

     The extension of the above procedure to the case of a linear defect
     \cite{Xia} consisting of $n$ broken bonds is straight forward as
     each additional  defect introduces into $h$ a $2 \times 2$ diagonal
     block of the form  given in  Eq.(\ref{hform}).  In this  case,  the
     series (Eq.(\ref{taylor})) cannot be summed analytically due to the
     presence  of  cross  terms.  An  analytic  summation  is,  however,
     unnecessary.  It is sufficient to note that
     \begin{equation}
     \label{Aform1}
     A_{ij}=x_i^1 ( \delta_{j,k_1} - \delta_{j,k_1+1} )
      + x_i^2 ( \delta_{j,k_2} - \delta_{j,k_2+1} )
     + \cdots x_i^n ( \delta_{j,k_n} - \delta_{j,k_n+1} )
     \end{equation}
     for bonds missing  between $k_i$ and $k_{i+1} \; (i=1,2 \cdots n)$,
     and  that  consequently  $A^q$  has the  same  form as  above.  The
     numbers $x_i^p$'s  occurring in $A^q$ are of course  $q$--dependent.
     It  follows  from this that the series can once again be written in
     the form
     \begin{equation}
     {(I + A)}^{-1} = I + R
     \end{equation}
     where   $R$  is  a   constant   matrix   of  the  form   given   in
     Eq.(\ref{Aform1}).  Clearly the action of $R$ on a vector of length
     $N$ can be  determined in $nN$  multiplicative  steps.  The overall
     complexity  of the algorithm in this case is thus ${\cal O} (N\ln N
     + n N )$.  Since the number of broken  bonds  required  to  observe
     various breakdown phenomena \cite{Xia} is much smaller than $N$,
     this results in a very large saving in the time required to perform
     the computations.

     It is interesting to note that the Green's  function for the broken
     bond system $G^{bb}$ satisfies the Dyson's equation \cite{Inkson}

     \begin{equation}
     G^{bb} = \tilde{G} + \tilde{G} (-h) G^{bb}
     \end{equation}

     The  matrix  $(-h)$ is thus the analog of the  potential  due to an
     impurity in an otherwise perfect lattice and the case of the single
     defect is the analog of the $\delta$  function  impurity at a given
     site in the system.

     The procedure developed above is for the situation where no pair of
     broken  bonds has any site in common.  This  ensures that the block
     matrices  introduced  in $h$ are always  $2\times 2$ rather than $m
     \times m , \; m >2$ The general case can be dealt with in a similar
     manner albeit with an increase in complexity.  Lastly, bonds can be
     added in the interior of the network  rather than being  eliminated
     since this does not violate the constraint $\sum_i d_i = 0$.

     \section{Summary And Discussion}
     \label{sec:conclu}

      To summarize, we have extended the Eikman--Himbergen  algorithm to
      a number of practically  occurring  situations.  In particular, we
      have  extended it to the case of a bus--bar  placed along one edge
      of a rectangular  array as also to that of electrically  connected
      bus--bars  placed along two parallel edges.  Finally we have shown
      that our  simulation  can be  carried  out, with  only a  marginal
      increase in complexity, for bonds  eliminated from or added to the
      array.  In all cases, the open edges can be  connected  to current
      sources having any profile whatsoever.

      The algorithms for bus--bars as also for perfect arrays subject to
      magnetic fields, noise, and/or current drives, all involve Green's
      functions  specific  to  the  lattice  in  question.  The  Green's
      function of interest can straightly--forwardly be constructed once
      the  eigen--basis  of the  corresponding  connectivity  matrix  or
      discrete  Laplacian has been  determined.  We have shown that this
      eigen--basis  consists  precisely of those linear  combinations of
      eigenfunctions  of the  connectivity  matrix for a {\it  periodic}
      lattice, which satisfy an appropriate set of boundary conditions.

      The addition and removal of bonds introduces additional boundaries
      often in the  interior  of the  array.  As a  result  our  earlier
      procedure  becomes  invalid.  An island of defects  created at the
      center of the array, e.g., defines a lattice  version of the sinai
      billiard  problem,which  classically  has only chaotic  solutions.
      However  if the  number  of bonds  eliminated  is  small,  Green's
      function  techniques  which  have been  extensively  used to study
      disordered solids can be very effectively  applied.  We have shown
      that this  techniques  yields an ${\cal O}(N\ln N+nN)$  algorithm.
      The extensions of such algorithms to the case of 3D arrays is easy
      since the corresponding  eigen--vectors are specified to be either
      $\exp  (i\veck  \cdot  \vecr )$ or $\cos  (k_x x )$  depending  on
      whether  the   boundaries   are   periodic   or  free   boundaries
      respectively.  The  eigen--values  are now changed to $ 8-2cos k_x
      -2 \cos k_y -2\cos k_z $.  In this case, one has to obviously  use
      3-D transforms.

      It is also  worth  noting  that  many of the  expressions  we have
      derived can be written in terms of continous functions, which are,
      of course, to be evaluated or sampled only at points  belonging to
      the array.  Now if we turn this  observation  around  and think of
      the discrete  lattice as being {\it  produced} by the  sampling of
      the continuous 2--D waveform, we make contact with a long-standing
      problem in electrical  engineering, viz.  that of the digitisation
      and subsequent  recovery of band-width limited analog  wave-forms.
      Usually such waveforms are processed as rectangular  spaced arrays
      i.\ e.\ they are periodically  sampled along the two  orthogonally
      independent   variables.  However,  one  can  also  use  hexagonal
      sampling  \cite{Merser}.  Once  discretised,  the waveforms can be
      processed as arrays of numbers  $x(n_1,n_2)$ by the computer where
      $(n_1, n_2)$ is a discrete  point on which the observable  $x$ has
      some value.  The  periodicity  of the observable  $x(n_1, n_2)$ is
      important as it specifies the  eigenfunctions  associated with the
      system.  The matrix $G_0^{-1}$ has precisely the information about
      the chosen  sampling  raster and the  periodicity  of the lattice.
      The raster  defines the total  number of nearest  neighbours  of a
      general site $(n_1, n_2)$ while the periodicity (or finiteness) of
      the lattice decides the  connectivity of the sites at the boundary
      of the  network.  The  eigenfunctions  used in both  cases are the
      same.  Indeed, the analogy goes deeper and we can think of the JJA
      itself   as  a   latticized   version   of  the   continuum   case
      (superconducting  islands embedded in a normal  background much as
      the Abrikosov  lattice  consists of normal  regions  embedded in a
      superconducting  background).  Finally, since hexagonally  sampled
      waveforms  can be  recovered  in ${\cal  O}(N\ln  N)$ steps,  this
      analogy opens up the very  interesting  possibility of working out
      ${\cal O}(N\ln N)$ algorithm for a triangular JJA.  This and other
      related issues are currently under active investigation.

\bibliographystyle{prsty}

\clearpage
\centerline{{\bf Figure Caption}}
\begin{itemize}
\item[Fig.1] The  figure  shows the  geometry  with  currents  $I$ being
             injected   and   extracted   at  $x=x_0$  and   $x=x_{L_x}$
             respectively.  The $\bullet$ represents the superconducting
             islands and the line joining them represents a junction.

\item[Fig.2] The figure shows the single bus--bar geometry with currents
             $I$ being injected $x=x_0$.  The short is at $x=x_{L_x+1}$.
             Note that one can use any current profile in this case.

\item[Fig.3] The figure shows the double  bus--bar  geometry with shorts
             at $x=x_0$ and $x=x_{L_x+1}$.
\end{itemize}

\end{document}